\documentclass[aps,amssymb,amsfonts,twocolumn,showpacs]{revtex4}
\usepackage{graphicx} \usepackage{bm}
 \usepackage{amsmath}
 \usepackage{amssymb}
 \usepackage{latexsym}
 \usepackage{amsfonts}
 \usepackage{epsfig}

\newcommand\br{\mathbf{r}}

 \newcommand{\bea}{\begin{eqnarray}}
 \newcommand{\ea}{\end{eqnarray}}
 \newcommand{\eea}{\end{eqnarray}}
 
\newcommand{\edd}{\varepsilon_{\rm dd}}
 
\newcommand{\ltsimeq}{\raisebox{-0.6ex}{$\,\stackrel
        {\raisebox{-.2ex}{$\textstyle <$}}{\sim}\,$}}

\begin{document}
\title{Thomas-Fermi versus one- and two-dimensional regimes of a trapped dipolar Bose-Einstein condensate}
\author{N.G. Parker and D.H.J. O'Dell}
\affiliation{Department of Physics and Astronomy, McMaster
University, Hamilton, Ontario, L8S 4M1, Canada}

\begin{abstract}
We derive the criteria for the Thomas-Fermi regime of a dipolar
Bose-Einstein condensate in cigar, pancake and spherical geometries.
This also naturally gives the criteria for the mean-field one- and
two-dimensional regimes. Our predictions, including the Thomas-Fermi
density profiles, are shown to be in excellent agreement with
numerical solutions. Importantly, the anisotropy of the interactions
has a profound effect on the Thomas-Fermi/low-dimensional criteria.
\end{abstract}


\maketitle
\paragraph*{Introduction.}
A Bose-Einstein condensate (BEC) is said to be in the Thomas-Fermi
(TF) regime when the interaction energy dominates the zero-point
energy \cite{TF_BEC,p+sbook}.  Despite being a dilute gas, in this regime the BEC is strongly affected by interactions and behaves in a highly non-ideal manner.  Furthermore, the TF description is relatively simple  compared to full Gross-Pitaevskii theory and facilitates exact results \cite{p+sbook}.
Consider a BEC containing $N$ atoms of mass $m$ confined to a
spherical harmonic trap $V(r)=m \omega^{2}r^{2}/2$, with
harmonic oscillator (HO) length
$l_{\mathrm{ho}}=\sqrt{\hbar/(m \omega)}$, and repulsive $s$-wave
interactions with positive scattering length $a$. The
TF regime occurs for large values of the parameter $N
a/l_{\mathrm{ho}}$, and thus typically arises for dense strongly
repulsive BECs.

Recently, a {\em dipolar BEC} was created with atomic dipoles
polarized in a common direction by an external field
\cite{griesmaier}. Far from any scattering resonances the interaction can be represented by a pseudo-potential which
includes the bare dipole-dipole interaction \cite{Yi,Bohn},
\begin{equation}
U(\br)= g\delta({\bf r})+\frac{C_{\mathrm{dd}}}{4 \pi}\, \hat{{\rm
e}}_{i} \hat{{\rm e}}_{j} \frac{\left(\delta_{i j}- 3 \hat{r}_{i}
\hat{r}_{j}\right)}{r^{3}}. \label{eqn:U}
\end{equation}
The first term describes {\it s}-wave scattering arising primarily
from van der Waals interactions, where $g=4\pi\hbar^2 a/m$. The second term describes the long-range part of the
interaction between dipoles aligned along the unit vector
$\hat{\mathbf{e}}$, where $C_{\rm dd}$ is the dipolar coupling
strength. An important quantity is the ratio between the dipolar and
{\it s}-wave interactions $\varepsilon_{\rm dd}=C_{\rm dd}/3g$
\cite{giovanazzi02}. For magnetic dipoles with moment $d$,  $C_{\rm dd}=\mu_0 d^2$, where $\mu_0$ is the
permeability of free space. The signs and magnitudes of $C_{\rm
dd}$ and $g$ can be controlled by rotation of the polarization axis
\cite{giovanazzi02}, and by a Feshbach resonance \cite{Werner05},
respectively. A large range of $\edd$ is therefore accessible and
has already begun
to be probed experimentally \cite{Werner05,koch,lahaye}. 

In contrast to the van der Waals interaction, the dipolar
interaction is long-range and anisotropic. This has dramatic
consequences upon the behavior of a BEC and can even lead to
collapse when $N$ and/or $C_{\rm dd}$ exceed critical values
\cite{goral,Santos00,Yi,odell,koch}.
Many properties of a dipolar BEC in the TF regime have already been
discussed, e.g., its density profile \cite{odell}, expansion
dynamics \cite{stuhler}, excitation frequencies
\cite{odell,giovanazzi2007}, rotation \cite{rickPRL} and vortices
\cite{odell07}. However, a systematic discussion of the criteria for the
Thomas-Fermi regime of a dipolar BEC is currently lacking. In this
paper we derive these criteria for the 
important cases of cigar-, pancake-, and spherically-shaped ground states. 
Our approach also
reveals the criteria for the mean-field 1D and 2D regimes. 

\paragraph*{Theory.}
At zero temperature the mean-field condensate wave function $\psi
\equiv \psi({\bf r},t)$ obeys the Gross-Pitaevskii equation (GPE)
\cite{p+sbook}. Stationary solutions, for which $\psi$ is real,
satisfy the time-independent GPE,
\begin{equation}
\left[-(\hbar^2/2m)\nabla^2+V({\bf r})+g\psi^2+\Phi_{\rm dd}({\bf
r},t)\right]\psi=\mu \psi, \label{eqn:GPE}
\end{equation}
where $\mu$ is the chemical potential and the atomic density $n({\bf
r})=\psi({\bf r})^2$ is normalised via $\int n({\bf r}) d^3{\bf
r}=N$.

We assume that confinement of the gas is provided by a
cylindrically-symmetric harmonic trap $V({\bf
r})=\frac{1}{2}m(\omega_x^2 \rho^2+\omega_z^2 z^2)$, where $\omega_x$
and $\omega_z$ are the radial and axial trap frequencies, and $\rho=\sqrt{x^2+y^2}$. The {\it
s}-wave interactions introduce a {\em local} mean-field potential
$g\psi^2$ while the dipolar interactions introduce a {\em non-local}
potential $\Phi_{\rm dd}$ given by \cite{Yi},
\begin{equation}
\Phi_{\rm dd}({\bf r})=\int d^3{\bf r}' U_{\rm dd}({\bf r}-{\bf
r}')\psi({\bf r}')^2, \label{eqn:phidd1}
\end{equation}
where $U_{\rm dd}({\bf r})$ is the second term in Eq.~(\ref{eqn:U}).

In a time-dependent situation, where $\psi$ is complex, kinetic
energy arises from both zero-point motion (density
gradients) and velocities (phase gradients). However,
in Eq.~(\ref{eqn:GPE}), only zero-point energy contributes. When
this is negligible we enter the TF regime and Eq.~(\ref{eqn:GPE})
becomes,
\begin{equation}
 gn({\bf r})+m(\omega_x^2 \rho^2+\omega_z^2 z^2)/2+\Phi_{\rm dd}({\bf r})=\mu. \label{eqn:TF}
\end{equation}
Equation~(\ref{eqn:TF}) is satisfied by
the well-known inverted parabola solution even in the presence of dipolar interactions \cite{odell}. This has the form,
\begin{equation}
 n({\bf r})=n_0\left[1-(\rho/R_x)^2-(z/R_z)^2 \right],
\label{eqn:parabola}
\end{equation}
where $n_0=15N/(8\pi R_z R_x^2)$ is the peak density.  The density is
zero beyond the BEC boundary which is specified by the TF radii, $R_x$ and $R_z$.  Expressions for $R_x$ and $R_z$ are presented in \cite{odell}.  For purely {\it s}-wave interactions, stable TF solutions only exist for repulsive interactions ($g>0$); under
attractive interactions ($g<0$), the zero-point kinetic energy is
crucial to stabilise the BEC against collapse. Additionally, the aspect ratio  $\kappa=R_x/R_z$ equals the trap ratio $\gamma=\omega_z/\omega_x$, while for dipolar interactions this is not generally true.

It is useful to introduce a fictitious electrostatic potential
$\phi({\bf r})=\int d {\bf
 r}' n({\bf r})/(4\pi |{\bf r} - {\bf r}'|)$.  This satisfies Poisson's equation $\nabla^2\phi=-n({\bf
 r})$.  For dipoles aligned in the $z$-direction, $\Phi_{\rm dd}$ can then
be expressed as \cite{odell},
 \begin{equation}
 \Phi_{\rm dd}({\bf r}) = -g \edd \left[3 \partial_z^2 \phi({\bf r}) +n({\bf r})\right]. \label{eq:dip}
 \end{equation}
Here the first term is anisotropic and long-range, while the second
is short-range and contact-like. Note that the dipolar potential inside the inverted parabola (\ref{eqn:parabola}) is \cite{odell},
 \begin{equation}
  \Phi_{\rm dd}=\frac{n_0 g \edd}{R_z^2}\left[\frac{\rho^2}{\kappa^2}-2z^2- f(\kappa)\left(R_z^2-\frac{3}{2}\frac{\rho^2-2z^2}{\kappa^2-1}\right) \right], \label{eqn:sph_Phidd2}
 \end{equation}
where $f(\kappa)=(1+2\kappa^2)/(1-\kappa^2)-3\kappa^2{\rm atanh}\sqrt{1-\kappa^2}/(1-\kappa^2)^{3/2}$ lies in the range $1 \geq f(\kappa)\geq -2$ and $\kappa$ is determined by a transcendental equation \cite{Yi,odell},
\begin{equation}
  3\kappa^2\edd\left[\left(\frac{\gamma^2}{2}+1\right)\frac{f(\kappa)}{1-\kappa^2}-1\right]=(1-\edd)(\kappa^2-\gamma^2).\label{eqn:trans}
\end{equation}

\paragraph*{Cigar: 3D TF vs 1D mean-field.}
Menotti and Stringari \cite{menotti} analysed the crossover between the 3D TF and 1D
mean-field regimes of a cigar-shaped {\it s}-wave BEC. We now
extend their approach to include dipolar interactions. We begin by
neglecting the axial trapping ($\omega_z=0$) and consider the BEC to
be uniform along $z$ with 1D density (number of atoms per unit
length) $n_1$. Then Eq.~(\ref{eq:dip})
reduces to the contact-like form,
\begin{equation}
  \Phi_{\rm dd}(\br)=-g\edd n(\rho). \label{eq:cigarphi}
\end{equation}
We can thus define an effective {\it s}-wave scattering length for the cigar $a_{\rm c}=a(1-\edd)$. The
dipoles in the cigar lie predominantly end-to-end
such that for $C_{\rm dd} >0$ ($C_{\rm dd}<0$) the net
dipolar interaction is attractive (repulsive).  

Introducing the dimensionless quantities $\rho'=\rho/l_x$ and $\psi_\rho(\rho')=l_x \psi(
\rho)/\sqrt{n_1}$, where
$l_x=\sqrt{\hbar/m\omega_x}$ is the radial HO length, Eq.~(\ref{eqn:GPE}) becomes,
\begin{equation}
\left(-\frac{\partial_{\rho'}^2}{2}-\frac{\partial_{\rho'}}{2\rho'}+\frac{\rho'^2}{2}+4
\pi a_{\rm c} n_1 {\psi_\rho}^2 \right)\psi_\rho
=\frac{\mu_1[a_{\rm c} n_1]}{\hbar \omega_x} \psi_\rho. \label{eq:cigargpe}
\end{equation}
The chemical potential $\mu_1[a_{\rm c} n_1]$ is a function of $n_1$: one can numerically tabulate this equation of state by solving (\ref{eq:cigargpe}) for different values of $n_{1}$.
The kinetic terms in Eq.~(\ref{eq:cigargpe}) become negligible in comparison to
the interaction terms when $a_{\rm c} n_1\gg 1$, in which
case the TF radial density profile has the form,
\begin{equation}
{\psi_\rho}^2=\frac{1}{4\pi a_{\rm c} n_1
}\left(\frac{\mu_1[a_{\rm c} n_1]}{\hbar\omega_x}-\frac{1}{2}\rho'^2
\right), \label{eq:TFdensity1}
\end{equation}
with TF radius $R_x=\sqrt{2 \mu_1[a_{\rm c} n_1] / \hbar \omega}$.
Use of the normalisation condition $\int {\psi_\rho}^2 2\pi \rho' d\rho'=1$
leads to the relation,
\begin{equation}
  \mu_1[a_{\rm c} n_1]/\hbar\omega_x=2\sqrt{a_{\rm c} n_1}. \label{eq:mu_TF}
\end{equation}

In the opposite regime of $a_{\rm c} n_1\ll 1$, termed the 1D
mean-field regime \cite{p+sbook}, the solution of
Eq.~(\ref{eq:cigargpe}) approximates the non-interacting radial HO
state $\psi_\rho=\pi^{-1/2} e^{-\rho'^2/2}$. Inserting this form into
Eq.~(\ref{eqn:GPE}), we obtain the chemical potential perturbatively
to be,
\begin{eqnarray}
\mu_1[a_{\rm c} n_1]/\hbar\omega_x=1+2a_{\rm c}n_1.\label{eq:mu_nonint}
\end{eqnarray}

We now wish to find the effect of finite axial trapping $\omega_z
\neq 0$, for which $n_1(z)$ becomes inhomogeneous.  We can easily estimate the correction to Eq.~(\ref{eq:cigarphi}) using the TF result.  Expansion of Eq.~(\ref{eqn:sph_Phidd2}) as $\kappa \rightarrow 0$ \cite{cigar_expansions} reveals that the leading correction  is of order
$\kappa^2$ and becomes negligible for $\kappa\ll 1$.  Thus if the variation
is sufficiently weak along $z$, Eq.~(\ref{eq:cigargpe}) still
defines the
{\em local} chemical potential along $z$, $\mu_1[a_{\rm c} n_1(z)]$, and we can
also employ the local density approximation,
\begin{equation}
 \mu_1[a_{\rm c} n_1(z)]+m\omega_z^2 z^2/2=\mu, \label{eq:lda}
\end{equation}
where $\mu$ is the {\em global} chemical potential of
Eq~(\ref{eqn:GPE}). At the axial boundary $R_z$, this gives
$\mu_1[a_{\rm c} n_1=0]+\frac{1}{2}m\omega_z^2 R_z^2=\mu$. Then, by defining
the function $\tilde{\mu}_1=(\mu_1[a_{\rm c}n_1(z)]-\mu_1[a_{\rm c} n_1=0])/\hbar
\omega_x$, we can rewrite Eq.~(\ref{eq:lda}) as \cite{menotti},
\begin{equation}
 \tilde{\mu}_1[a_{\rm c} n_1(z)]=(\alpha^{2}/2)\left[1-(z/R_z)^2 \right] \label{eq:mu1}
\end{equation}
where $\alpha=l_x R_{z}/l_z^{2}$. We can determine $n_{1}(z)$ by
using the equation of state $\mu_{1}[a_{\rm c} n_{1}]$ to invert Eq.\
(\ref{eq:mu1}), i.e.,
$n_1(z)=\tilde{\mu}_1^{-1}[\tilde{\mu}_1[a_{\rm c} n_1(z)]]/a_{\rm c}$. For each choice
of $\alpha$ we still need to know $R_{z}$, and this is obtained from the
normalization condition $\int n_1(z)dz=N$ which can be written as,
\begin{equation}
 \alpha \int^{1}_{-1}\tilde{\mu}_1^{-1} \left[\frac{1}{2}\alpha^2(1-\zeta^2) \right]d\zeta=Na(1-\edd)\frac{l_x}{l_z^2},\label{eq:cigar}
\end{equation}
where $\zeta=z/R_z$. On the right hand side we identify a natural
dimensionless parameter. We enter the radial TF regime (and
therefore the full 3D TF regime) when,
\begin{equation}
 Na (1-\edd)l_x/l_z^2 \gg 1. \quad \quad \mbox{3D TF cigar} \label{eq:cigarTF}
\end{equation}
In this case we can use Eq.~(\ref{eq:mu_TF}) to give the density profile,
\begin{equation}
 n_1(z)=\frac{1}{a(1-\edd)}\left(\frac{l_x R_z }{2 l_z^2} \right)^{4}\left(1-\frac{z^2}{R_z^2} \right)^2 \label{eq:density_3Dcigar},
\end{equation}
where $R_z=[15Na(1-\edd)l_z^8/l_x^4]^{1/5}$. When $Na (1-\edd)l_x/l_z^2 \ll1$ we enter the 1D mean-field
regime. Then, Eq.~(\ref{eq:mu_nonint}) leads to the density profile,
\begin{equation}
  n_1(z)=\frac{1}{a(1-\edd)}\left(\frac{l_x R_z}{2l_z^2} \right)^{2}\left(1-\frac{z^2}{R_z^2} \right) \label{eq:density_1Dmeanfield},
\end{equation}
where $R_z=[3Na(1-\edd)l_z^2/l_x^2 ]^{1/3}$.  Importantly, at $\edd=1$ the overall contact interactions vanish
and the TF cigar criterion (\ref{eq:cigarTF}) cannot be satisfied.
When $\edd=0$ we retrieve the results of \cite{menotti}.

\begin{figure}[t]
\centering
\includegraphics[width=8.5cm,clip=true]{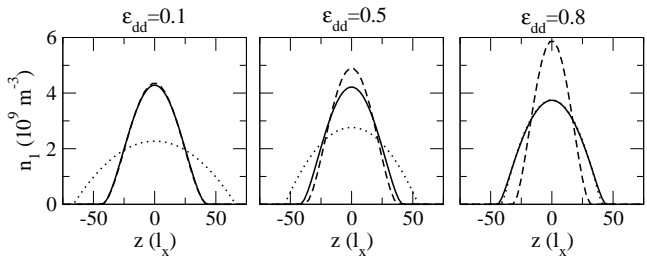}
\caption{Axial density $n_1(z)$ of a cigar BEC for various values of
$\edd$ according to the GPE (solid lines), 3D TF cigar prediction of
Eq.~(\ref{eq:density_3Dcigar}) (dashed line) and 1D mean field
prediction of Eq.~(\ref{eq:density_1Dmeanfield}) (dotted line). We
employ $Na l_x/l_z^2=100$, $\gamma=0.1$, $\omega_x=2\pi \times
50$Hz and $^{52}$Cr atoms.} \label{fig:cigar}
\end{figure}

We now illustrate the key dependence on $\edd$. We have obtained
numerically the ground states of the 3D dipolar GPE, following
the procedure outlined in \cite{goral02}. The axial density profiles
for a cigar BEC are compared in Fig.~(\ref{fig:cigar}) for various
values of $\edd$, according to the GPE (solid line), TF prediction
(dashed line) and 1D mean-field prediction (dotted line). For weak
dipolar interactions $\edd=0.1$, giving $Na (1-\edd)l_x/l_z^2 \gg 1$,
the GPE profile is in excellent agreement with the 3D TF cigar
prediction. For strong dipolar interactions $\edd=0.8$, the 1D
mean-field regime becomes an excellent description of the profile.
For intermediate dipolar interactions $\edd=0.5$ neither 
analytic form agrees well with the GPE results.  In \cite{stuhler} observations of the density and expansion 
of a cigar dipolar BEC were compared with TF predictions. For their
parameters, including $\edd\approx 0.16$ and $N\approx
5\times 10^5$, we obtain $N a(1-\edd)l_x/l_z^2 \approx 50
\gg 1$ thus satisfying the cigar TF criteria.

\paragraph*{Pancake: 3D TF vs 2D mean-field.}
We now consider a highly flattened pancake BEC and follow a similar
methodology to the cigar BEC. We initially assume that the BEC has
infinite radial extent and uniform 2D density $n_2$. Then, Poisson's
equation $\nabla^2 \phi=-n({\bf r})$ reduces to $\partial_z^2
\phi=-n(z)$, such that,
\begin{equation}
  \Phi_{\rm dd}(\br)=2g\edd n(z), \label{eq:pancakephi}
\end{equation}
This is, again, contact-like, giving an effective scattering length $a_{\rm p}=(1+2\edd)a$. In the pancake the dipoles are predominantly
side-by-side and so the net interaction is repulsive (attractive)
for $C_{\rm dd}>0$ ($C_{\rm dd}<0$).  Introducing the dimensionless parameters $z'=z/l_z$ and $\psi_z(z')=\sqrt{l_z/n_2}\psi({\bf r})$, where $l_z=\sqrt{\hbar/m\omega_z}$ is the axial HO length, the GPE
becomes,
\begin{equation}
\left(-\frac{\partial_{z'}^2}{2}+\frac{z'^2}{2}+4 \pi a_{\rm p} l_z
n_2 \psi_z^2 \right) \psi_z
=\frac{\mu_2[a_{\rm p} l_z n_2]}{\hbar \omega_z} \psi_z.
\end{equation}
The axial TF regime exists when $a_{\rm p} l_z n_2 \gg 1$. By
normalising the corresponding TF density profile via $\int \psi_z^2
dz'=1$, we obtain the TF chemical potential,
\begin{equation}
 \mu_2[a_{\rm p} l_z n_2]/\hbar \omega_z=\left(6 \pi a_{\rm p} l_z n_2 \right)^{2/3}/2. \label{eq:pancakeTF}
\end{equation}
In the opposite regime of $a_{\rm p} l_z n_2 \ll 1$, we
enter the 2D mean-field regime. Here $\psi_z$ approximates the
ground HO state $\psi_z=\pi^{-1/4}e^{-z'^2/2}$ and the chemical
potential is found perturbatively to be,
\begin{equation}
 \mu_2[a_{\rm p} l_z n_2]/\hbar \omega_z=1+2\sqrt{2\pi}a_{\rm p} l_z n_2. \label{eq:pancake2D}
\end{equation}

We now consider introduce finite radial trapping $\omega_r\neq0$. Expanding Eq.~(\ref{eqn:sph_Phidd2}) for $\kappa \rightarrow \infty$ \cite{pancake_expansions} shows that 
the leading correction to Eq.~(\ref{eq:pancakephi}) is of the order
$\kappa^{-1}$ and negligible when $\kappa\gg1$.  Defining the function $\tilde{\mu}_2=(\mu_2[a_{\rm p} l_z n_2(z)]-\mu_2[a_{\rm p} l_z n_2=0])/\hbar
\omega_z$ and employing the normalisation condition $\int n_2(\rho)2 \pi \rho d\rho=N$ we
obtain,
\begin{equation}
 2\pi \beta^{2}\int^{1}_{0}\tilde{\mu}^{-1} \left[\frac{\beta^{2}}{2} (1-s^2) \right]s~ds=\frac{N a (1+2\edd)l_z^{3}}{l_x^4},\label{eq:pancake}
\end{equation}
where $s=\rho/R_x$ and $\beta=l_z R_x/l_x^{2}$. We therefore
enter the axial TF regime, and hence the 3D TF regime, when,
\begin{equation}
 Na (1+2\edd)l_z^{3}/l_x^{4} \gg 1. \quad \mbox{3D TF pancake} \label{eq:TFpancakeCondition}
\end{equation}
For $\edd=0$ we obtain the same result as in \cite{delgado}. 
The corresponding 2D TF density profile is,
\begin{equation}
 n_2(r)=\frac{1}{3\pi l_z a(1+2\edd)}\left(\frac{l_z R_x }{l_x^2} \right)^{3}\left(1-\frac{\rho^2}{R_x^2} \right)^{3/2} \label{eq:density_3Dpancake},
\end{equation}
where $R_x=[15N a (1+2\edd)l_x^6/l_z^2]^{1/5}$.

In the 2D mean-field regime of $Na(1+2\edd)l_z^{3}/l_x^{4} \ll 1$ we obtain,
\begin{equation}
  n_2(r)=\frac{1}{\sqrt{2\pi} l_z a(1+2\edd)}\left(\frac{l_z R_x}{2l_x^2} \right)^{2}\left(1-\frac{\rho^2}{R_x^2} \right) \label{eq:density_2Dmeanfield},
\end{equation}
where $R_x=[16Na(1+2\edd)l_x^4/\sqrt{2\pi}l_z]^{1/4}$.

The experiment \cite{koch} featured a pancake BEC with $\gamma=10$.
For their parameters ($\omega_x=2\pi \times 330$, $N=25,000$ and
$\edd\approx0.16$) we find $N a(1+2\edd)l_z^3/l_x^4=7$,
suggesting that the initial BEC was not in the TF regime.

\begin{figure}[t]
\centering
\includegraphics[width=8.5cm,clip=true]{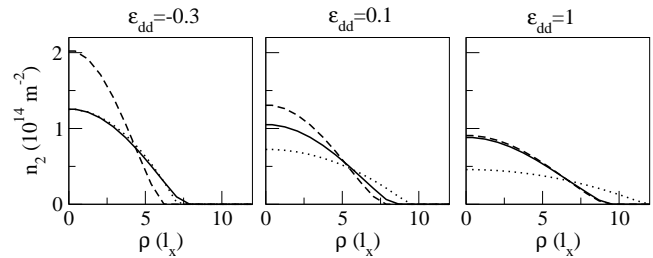}
\caption{2D density $n_2(\rho)$ of a pancake BEC for various values of
$\edd$ according to the GPE (solid lines), TF pancake prediction of
Eq.~(\ref{eq:density_3Dpancake}) (dashed line) and 2D mean field
prediction of Eq.~(\ref{eq:density_2Dmeanfield}) (dotted line). We
employ $Na l_z^3/l_x^4=10$, $\gamma=10$, $\omega_x=2\pi
\times 5$Hz and $^{52}$Cr atoms.}\label{fig:pancake}
\end{figure}
In a very flat pancake, the net contact interactions become zero for
$\edd=-0.5$ and the TF regime cannot be supported. We demonstrate
the role of $\edd$ in Fig.~\ref{fig:pancake} which presents the 2D
density profile of a pancake BEC. For $\edd=-0.3$, the net
interactions are very weak and the 2D mean-field prediction (dotted
line) agrees well with the GPE profile (solid line). Conversely, for
$\edd=1$ the combined interactions are relatively strong giving
excellent agreement with the 3D TF prediction (dashed line). For an
intermediate value of $\edd=0.1$, the exact solution lies in between
the two analytic predictions.

\paragraph*{Approximately spherical BEC.}
Away from the cigar and pancake limits $\Phi_{\rm dd}$ no longer
reduces to a local interaction and we must consider its full form.
In order to proceed we shall assume the validity of the 3D TF
solution (\ref{eqn:parabola}) and use energetic arguments to check
its self-consistency. The combined {\it s}-wave and dipolar
interaction energy for the TF solution is \cite{odell},
\begin{eqnarray}
E_{\rm i}=\int \left(\frac{g}{2}\psi^4 +\Phi_{\rm dd}
\psi^2\right)d^3{\bf r}=\frac{15N^2 g \kappa}{28\pi R_x^3}\left[
1-f(\kappa)\edd \right]\label{eq:dipE}
\end{eqnarray}
We observe that when $\edd=1/f(\kappa)$ the net interactions vanish and the TF regime cannot be achieved.  The kinetic energy of a spherical TF distribution is
$E_{\rm k}\approx 5N\hbar^2/2mR_x^2$, up to a logarithmic correction \cite{p+sbook}. The TF approximation is valid when $E_{\rm i}/E_{\rm k}\gg1$ which, using the above expressions and dropping the numerical prefactors ($\approx 1$), becomes $(N a\kappa/R_x)\left[1-f(\kappa)\edd \right] \gg 1$.  From Eq.~(\ref{eqn:trans}) we find that a spherical ($\kappa=1$) BEC occurs when $\edd=\edd^0$ where $\edd^0= (5/2)(\gamma^2-1)(\gamma^2+2)$.  Introducing the parameter $\delta=\edd-\edd^0$ and expanding Eq.~(\ref{eqn:trans}) to first order in $\delta$ and $(\kappa-1)$, we obtain the relation $\kappa=1-(7\delta/15)(\gamma^2+2)^2/(12\gamma^2-2\gamma^4-3)$.  Furthermore, we expand the expression for $R_x$ in \cite{odell} and $f(\kappa)$ \cite{spherical_expansions}.  To first order in $\delta$, the TF criterion for an approximately spherical dipolar BEC is then,
\begin{equation}
\left(\frac{Na}{a_x} \right)^{4/5} \left(\frac{2+\gamma^2}{45} \right)^{1/5}\left[1-\frac{14\gamma^2(2+\gamma^2)\delta}{15(12\gamma^2-3-2\gamma^4)} \right]\gg 1. \label{eqn:sph_condition}
\end{equation}
Since we employ $R_x$ as the length scale rather than $a_x$, this criteria has a different form from the usually-quoted $Na/a_x$.  
We have confirmed that this gives a very good approximation for small deviations, typically up to $10\%$, from $\kappa=1$.  It is important to note that Eq.~(\ref{eqn:trans}) has both stable and unstable static solutions \cite{odell}.  Stable solutions of the full transcendental equation (\ref{eqn:trans}) for $\kappa=1$ exist only when $(12\gamma^2-3-2\gamma^4)>0$  or equivalently $0.5\ltsimeq\gamma\ltsimeq2.4$. Hence the factor multiplying $\delta$ in Eq.~(\ref{eqn:sph_condition}) is always positive and finite for the cases of interest.

Equation~(\ref{eqn:sph_condition}) reveals the sensitivity of the interactions to deviations from a perfectly spherical shape, as characterised by $\delta$. For a perfectly spherical BEC ($\delta=0$), the dipolar energy is zero and the TF
criterion (\ref{eqn:sph_condition}) reduces to its {\it s}-wave form.  $\Phi_{\rm dd}$ itself does not vanish but takes on the saddle-shaped form $\Phi_{\rm dd}=2n_0 g
\edd(\rho^2-2z^2)/5R_z^2$ as obtained by taking the limit $\kappa \rightarrow 1$ of  Eq.~(\ref{eqn:sph_Phidd2}) \cite{spherical_expansions}.  For
$\delta>0$ ($\delta<0$), the BEC is slightly elongated (flattened) and the net dipolar interactions are attractive (repulsive). 
In \cite{lahaye}, an approximately
spherical dipolar BEC was created ($\omega_x \approx 2\pi \times
480$Hz, $N=30,000$, $\kappa \approx 1$ and $a\approx 5 $nm) for which the left side of Eq.~(\ref{eqn:sph_condition}) equals $45$, confirming that it was in the TF
regime.

The TF criteria given in this paper do not specify whether the putative ground state actually exists.  For a pancake dipolar BEC, radial density wave structures have been predicted for $\edd \rightarrow \infty$ \cite{ronen}, and so the assumption of homogeneous radial density leading to Eq.~(\ref{eq:pancakephi}) would not hold. Furthermore, a stable ground state may not exist; the Bogoliubov spectrum for a uniform dipolar BEC \cite{goral} predicts an instability to density fluctuations when $\edd$ is outside of the range $-1/2<\edd<1$ for $a>0$.   Although trapping can significantly extend this range of stability, it must be
mapped out by solving the Bogoliubov de Gennes equations numerically \cite{ronen}.  However, within this range of stability and away from density wave structures, the
TF regime is both stable and has the inverted parabola profile, and
so the criteria should hold.

\paragraph*{Conclusions.}
For cigar-shaped and pancake-shaped condensates, the non-trivial
dipolar interactions reduce to a simple contact-like form.  We
identify their TF criteria, Eqns~(\ref{eq:cigarTF}) and (\ref{eq:TFpancakeCondition}), which also determine the 1D and 2D mean-field regimes. For the cigar (pancake) the net interactions are
proportional to $1-\edd$ ($1+2\edd$) such that that it becomes
increasingly difficult to remain in the TF regime as
$\edd\rightarrow 1$ ($\edd \rightarrow -1/2$). This highlights the
profound effect of the anisotropic dipolar interactions. Our
predictions are in excellent agreement with full numerical solutions. Furthermore, for the more complicated case of a
spherical condensate we determine the self-consistency criterion (\ref{eqn:sph_condition}) for
Thomas-Fermi behaviour. These criteria are of relevance to current
theoretical and experimental studies of dipolar condensates.

We thank the Canadian Commonwealth Scholarship Program (NGP) and NSERC (DHJOD) for funding and R. M. W. van Bijnen for stimulating discussions.

\end{document}